\documentclass[a4paper,10pt]{article}
\usepackage{graphicx}
\usepackage{epstopdf}
\usepackage{amsfonts,amsmath,amssymb}
\usepackage{hyperref}

\usepackage{epsfig,multicol}

\newcommand\fverb{\setbox\pippobox=\hbox\bgroup\verb}
\newcommand\fverbit{\egroup\item[\fbox{\unhbox\pippobox}]}

\newbox\pippobox

\begin{document}
\title{\bf Cosmic acceleration via space-time-matter theory }
\author{Raziyeh Zaregonbadi\thanks{Electronic address: r.zaregonbadi@malayeru.ac.ir}\,\,
\\
\small Physics Department, Faculty of Science, Malayer University, Malayer, Iran}

\maketitle
\begin{abstract}
\noindent
We consider the space-time-matter theory (STM) in a five-dimensional vacuum space-time
with a generalized FLRW metric to investigate the late-time acceleration of the universe. 
For this purpose, we derive the four-dimensional induced field equations 
and obtain the evolution of the state parameter with respect to the redshift.
Then, we show that with consideration of the extra dimension scale factor to be a linear function of redshift, 
this leads to a model which gives an accelerating phase in the universe.
Moreover, we derive the geodesic deviation
equation in the STM theory
to study the relative acceleration of the parallel geodesics of this
space-time, and also, obtain the observer area-distance as a measurable quantity to compare this theory with two other models.
\end{abstract}

\bigskip
\section{Introduction}\label{sec1}
\indent

The measurements of type Ia supernovae luminosity distances
indicate that the universe is currently undergoing an accelerated
expansion phase~\cite{Riess98}-\cite{Riess04}. 
However, general relativity, as the standard theory of gravitation, cannot result in such an expanding phase.
In this direction, several attempts have been made to modify general relativity in a way to explain this expansion, see, e.g., Refs.~\cite{Carroll}-\cite{Wu}.
Among these attempts, which go all under the name of modified gravitational theories, 
in this study we focus on the space-time-matter theory (STM) which is based on the existence of an extra dimension.
This theory, which is a generalization of the Kaluza-Klein theory~\cite{Klein,Appelquist},  
suggests that the $ 5D $ field equations
without any source can be reduced to the $ 4D $  
ones with an induced matter source whose energy-momentum tensor is defined in terms of the extra part of the geometry\cite{Wesson92,Wesson99}.

In this study, we use the STM theory to investigate 
the late-time acceleration of the universe.
In fact, we derive the $ 4D $ induced field equations from 
the $ 5D $ vacuum ones to show that the extra dimension can appear 
as a source of the acceleration phase of the universe without the requirement of the mysterious dark energy.
Also, to make our investigations more instructive,
we study the motion of the time-like and null test particles through 
the geodesic deviation equation (GDE) of the STM theory and obtain the observer area-distance as a measurable physical quantity.
Then, to compare the obtained results with other modified gravitational theories, we choose the $ \Lambda CDM $ model (in which 
the accelerated expansion of the universe is explained through adding the cosmological constant term into the action) and a particular minimal coupling model of the $ f(R,T) $ gravity~\cite{Zare} (in which the acceleration phase is explained through coupling between the matter source, as a geometrical curvature induction of matter, and the pure geometry).

The present work is organized as follows: 
In Sect.~2, we review how the $ 4D $ induced Einstein tensor
is obtained from the $ 5D $ vacuum space-time in STM theory.
In Sect.~3, taking a generalization of the Friedmann-Lema\^{i}tre-Robertson-Walker (FLRW) metric, 
we derive the cosmological equations, and also obtain the state parameter with respect to the redshift.
In Sect.~4, we derive the GDE for the time-like and null geodesics, 
and the observer area-distance. 
Moreover, the obtained result is compared with the $ \Lambda CDM $ model and a minimal coupling model of the $ f(R,T) $ gravity. 
Finally, in Sect.~5, we present the conclusions.
\\

\section{Space-time-matter theory}\label{sec 2}
\indent
As it was mentioned above, the STM theory is a generalization of the Kaluza-Klein theory in which the existence of an extra dimension is considered as the origin of matter. To show how the $ 4D $ Einstein equations with a non-trivial energy-momentum tensor 
can be induced from a $ 5D $ vacuum space-time, we begin with the following $ 5D $ line element~\cite{Wesson92}
\begin{equation}\label{eq1}
d{S^2} = {g_{ab}}({x^c})d{x^a}d{x^b} = 
{g_{\mu \nu }}({x^c })d{x^\mu }d{x^\nu } +
 \varepsilon {\phi ^2}({x^c }){(d{x^4})^2},
\end{equation}
where the Latin indices run from 0 to 4, the Greek indices run from
0 to 3, and $ {x^{4}} $ is a coordinate associated with the $ 5th $ dimension. 
Also, $ \varepsilon  =  \pm 1 $ allows us to choose the extra dimension to be either
time-like or space-like, and $ \phi ({x^c}) $ is a scalar field that depends
on all coordinates. 
By inserting Eq.~(\ref{eq1}) into the well-known relation
\begin{equation}\label{eq2}
^{\left( 5 \right)}{R_{ab}} = {\left( {\Gamma _{ab}^c} \right)_{,c}} - 
{\left( {\Gamma _{ac}^c} \right)_{,b}} 
+ \Gamma _{ab}^c\Gamma _{cd}^d - 
\Gamma _{ad}^c\Gamma _{bc}^d=^{\left( 5 \right)}{R_{\mu \nu }}+^{\left( 5 \right)}{R_{44}},
\end{equation} 
the following relations 
\begin{equation}\label{eq3}
^{\left( 5 \right)}{R_{\mu \nu }}= {^{\left( 4 \right)}{R_{\mu \nu }}} - \frac{{{\phi _{\mu ;\nu }}}}{\phi } + \frac{\varepsilon }{{2{\phi ^2}}}\left[ {\frac{{\mathop \phi \limits^ *  \mathop {{g_{\mu \nu }}}\limits^ *  }}{\phi } - \mathop {{g_{\mu \nu }}}\limits^{ *  * }  + {g^{\alpha \beta }}\mathop {{g_{\mu \alpha }}}\limits^ *  \mathop {{g_{\nu \beta }}}\limits^ *   - \frac{1}{2}\left( {{g^{\alpha \beta }}\mathop {{g_{\alpha \beta }}}\limits^ *  \mathop {{g_{\mu \nu }}}\limits^ *  } \right)} \right],
\end{equation}
and
\begin{equation}\label{eq4}
^{\left( 5 \right)}{R_{44}} =  - \varepsilon \phi\,{\Box} \phi  - \frac{1}{4}\left( {\mathop {{g^{\mu \nu }}}\limits^ *  \mathop {{g_{\mu \nu }}}\limits^ *  } \right) - \frac{1}{2}\left( {{g^{\mu \nu }}\mathop {{g_{\mu \nu }}}\limits^{ *  * } } \right) + \frac{{\mathop \phi \limits^ *  }}{{2\phi }}\left( {{g^{\mu \nu }}\mathop {{g_{\mu \nu }}}\limits^ *  } \right),
\end{equation}
are obtained, where $ ^{\left( 4 \right)}{R_{\mu \nu }} $ is the $ 4D $ Ricci tensor, ${\phi _\mu } \equiv \partial \phi /\partial {x^\mu }$, $ {\Box} \equiv {\nabla _\mu }{\nabla ^\mu } $, and the partial derivative with respect to the extra coordinate is denoted by $ * $. 
Now, taking the $5D$ space-time to be a vacuum space-time (i.e., $ {}^{\left( 5 \right)}{G_{ab}} = 0 $ which leads to the conditions  ${}^{\left( 5 \right)}{R_{\mu \nu }} = 0$ and ${}^{\left( 5 \right)}{R_{44}} = 0$) Eq. (\ref{eq3}) gives
the $ 4D $ Ricci tensor as
\begin{equation}\label{n5}
{^{\left( 4 \right)}{R_{\mu \nu }}}=\frac{{{\phi _{\mu ;\nu }}}}{\phi } - \frac{\varepsilon }{{2{\phi ^2}}}\left[ {\frac{{\mathop \phi \limits^ *  \mathop {{g_{\mu \nu }}}\limits^ *  }}{\phi } - \mathop {{g_{\mu \nu }}}\limits^{ *  * }  + {g^{\alpha \beta }}\mathop {{g_{\mu \alpha }}}\limits^ *  \mathop {{g_{\nu \beta }}}\limits^ *   - \frac{1}{2}\left( {{g^{\alpha \beta }}\mathop {{g_{\alpha \beta }}}\limits^ *  \mathop {{g_{\mu \nu }}}\limits^ *  } \right)} \right],
\end{equation}
and the $4D$ Ricci scalar as
\begin{equation}\label{eq6}
{^{\left( 4 \right)}R} = {^{\left( 4 \right)}{R_{\mu \nu }}}~{g^{\mu \nu }} = 
\frac{{\Box\phi }}{\phi } - \frac{\varepsilon {g^{\mu \nu }}}{{2{\phi ^2}}}\left[ {\frac{{\mathop \phi \limits^ *  \mathop {{g_{\mu \nu }}}\limits^ *  }}{\phi } - \mathop {{g_{\mu \nu }}}\limits^{ *  * }  + {g^{\alpha \beta }}\mathop {{g_{\mu \alpha }}}\limits^ *  \mathop {{g_{\nu \beta }}}\limits^ *   - \frac{1}{2}\left( {{g^{\alpha \beta }}\mathop {{g_{\alpha \beta }}}\limits^ *  \mathop {{g_{\mu \nu }}}\limits^ *  } \right)} \right].
\end{equation}
Hence, the Einstein tensor in the $ 4D $ space-time will be as follows
\begin{equation}\label{eq7}
{^{\left( 4 \right)}G_{\mu \nu }}= {^{\left( 4 \right)}R_{\mu \nu }}- \frac{1}{2}{{g_{\mu \nu }}}{^{\left( 4 \right)}R}\equiv {^{(4)}T_{\mu \nu }^{[\rm IM]}},
\end{equation}
where $ ^{(4)}T_{\mu \nu }^{[\rm IM]} $ is the energy-momentum tensor of the geometrical induced matter in the $4D$ space-time (which was introduced for the first time in the Wesson's induced matter theory~\cite{Wesson92, Wesson99}).
Now, inserting Eqs.~(\ref{n5}) and (\ref{eq6}) into Eq.~(\ref{eq7}) gives the expression of $ ^{(4)}T_{\mu \nu }^{[\rm IM]} $ as
\begin{equation}\label{tind} 
{^{(4)}T_{\mu \nu }^{[\rm IM]}}=\frac{1}{\phi }\left( {{\phi _{\mu ;\nu }} - \frac{1}{2}{g_{\mu \nu }}\Box\phi } \right)
+\frac{\varepsilon }{{2{\phi ^2}}}\left[ {\frac{{\mathop \phi \limits^ *  \mathop {{g_{\mu \nu }}}\limits^ *  }}{\phi } - \mathop {{g_{\mu \nu }}}\limits^{ *  * }  + {g^{\alpha \beta }}\mathop {{g_{\mu \alpha }}}\limits^ *  \mathop {{g_{\nu \beta }}}\limits^ *   - \frac{1}{2}\left( {{g^{\alpha \beta }}\mathop {{g_{\alpha \beta }}}\limits^ *  \mathop {{g_{\mu \nu }}}\limits^ *  } \right)} \right],
\end{equation}
which can be rewritten as 
\begin{equation}\label{tti}
{^{(4)}T_{\mu \nu }^{[\rm IM]}}=\frac{2}{\phi }\left( {{\phi _{\mu ;\nu }} - \frac{1}{4}{g_{\mu \nu }}\Box\phi } \right)- \frac{1}{4}{g_{\mu \nu }}{^{(4)}R}.
\end{equation}
Considering the above relation, the Bianchi identity ${\nabla ^\mu }\left({^{(4)}G_{\mu \nu }}\right) = 0 $ leads to the constraint
\begin{equation}\label{constraint}
\frac{{{\nabla ^\mu }\phi }}{\phi }\left( {{^{(4)}R_{\mu \nu }} - {^{(4)}G_{\mu \nu }}} \right) - \frac{^{(4)}R}{{4\phi }}{\nabla _\nu }\phi + \frac{3}{{4\phi }}{\nabla _\nu }\left( {\Box\phi } \right) - \frac{1}{4}{\nabla _\nu }(^{(4)}R )= 0,
\end{equation}
in the context of STM theory. We shall use this constraint for investigating a specific model in the next section.
\\

\section{Cosmological equations}\label{sec 3}
\indent

In order to study the evolution of the universe in the STM theory,
we consider the metric on the $ 4D $ manifold to be the spatially flat FLRW metric.
Hence, the $ 5D $  line element will be
\begin{equation}\label{eq8}
d{S^2} = 
f\left(x^4\right)\left[ - d{t^2} + {a^2}(t){\delta _{ij}}d{x^i}d{x^j} \right]+ {\phi ^2}(t){({dx^4})^2},
\end{equation}
where the extra dimension considered as a space-like one (i.e., in the metric (\ref{eq1}) we take $ \varepsilon $ equal to $1$), and also
$ f\left(x^4\right) $ is a function of the extra dimension.
We assume the $ 5D $ space-time to be a family of foliated hyper-surfaces 
where any possibility of $ x^4={x^4_0}$ defines a $ 4D $ space-time with
$f\left({x^4_0}\right)=const$. 
Without loss of generality, we set $f\left({x^4_0}\right)$ equal to 1. For the metric above, the constraint~(\ref{constraint}) 
will be of the form
\begin{equation}\label{conss}
R\dot \phi  - \dot R\phi  + 3{\partial _0}\left( {\Box\phi } \right) = 0.
\end{equation}

As an example, we consider the case of $ f\left(x^4\right)=(x^4/x^4_0)^{2n} $, in which the parameter $ n $ is a constant. In this case the constraint (\ref{conss}) leads to the relation
\begin{equation}
n\dot \phi \left( {4n - 1} \right) = 0,
\end{equation} 
where by considering $ \dot \phi $ and $n \ne 0  $, we obtain $ n=1/4 $. Thus, the Bianchi identity puts a restriction on the functionality of the metric with respect to the extra dimension.
Now, inserting the metric (\ref{eq8}) into the Eq. (\ref{eq7}) gives the following cosmological equation
\begin{equation}\label{eq12}
{H^2} = \frac{1}{{2\phi }}\left( {\ddot \phi  - H\dot \phi } \right) + \frac{^{(4)}R}{{12}},
\end{equation}
as the Friedmann like equation, and
\begin{equation}\label{eq13}
\frac{{\ddot a}}{a} =  - \frac{1}{{2\phi }}\left( {\ddot \phi  - H\dot \phi } \right) + \frac{^{(4)}R}{{12}},
\end{equation}
as the generalized Raychaudhuri equation, where the dot represents derivation with respect to the time and $ H\left( t \right) \equiv \dot a/a $ is the Hubble parameter. Then, straightforwardly we have
\begin{equation}\label{h dot}
\dot H = \frac{{\ddot a}}{a} - {H^2} = \frac{{H\dot \phi  - \ddot \phi }}{{\phi }}
\end{equation}
and 
\begin{equation}\label{R frid}
^{(4)}R = 6\left( {2{H^2} + \dot H} \right).
\end{equation}
Assuming the induced matter to be a perfect fluid, the energy density and the pressure of this fluid will be
\begin{equation}\label{ro}
{\rho ^{[\rm IM ]}} =  - {g^{00}}T_{00}^{[\rm IM]} = \frac{3}{{2\phi }}\left( {\ddot \phi  - H\dot \phi } \right) + \frac{^{(4)}R }{4},
\end{equation}
and
\begin{equation}\label{press}
{p^{[\rm IM ]}} = \frac{1}{3}{g^{ii}}T_{ii}^{[IM]} = \frac{1}{{2\phi }}\left( {\ddot \phi  - H\dot \phi } \right) - \frac{^{(4)}R }{4}.
\end{equation}
Having the above expressions for ${\rho ^{[\rm IM ]}}$ and ${p^{[\rm IM ]}}$, Eqs. (\ref{eq12}) and (\ref{eq13}) are rewritten respectively as 
\begin{equation}\label{fri1}
{H^2} = \frac{1}{3}{\rho ^{\left[ {\rm IM} \right]}},
\end{equation}
and
\begin{equation}\label{fri2}
\frac{{\ddot a}}{a} =  - \frac{1}{6}\left( {{\rho ^{\left[ {\rm IM} \right]}} + 3{p^{\left[ {\rm IM} \right]}}} \right).
\end{equation}
Now, to study the accelerating phase of the universe we need to know the evolution of the state parameter $ w $. 
For this purpose, we use the following definition
\begin{equation}\label{statp}
w(t)\equiv \frac{{{p^{[\rm IM]}}}}{{{\rho ^{[\rm IM]}}}} = - 1 - \frac{{2\dot H}}{{3{H^2}}}.
\end{equation}
Inserting Eqs.~(\ref{eq12}) and (\ref{h dot}) into (\ref{statp}), gives the expression for the state parameter as
\begin{equation}
w(t)= \frac{{2(\ddot \phi  - H\dot \phi ) - {^{(4)}R}~ \phi }}{{6\left( {\ddot \phi  - H\dot \phi } \right) + {^{(4)}R}~ \phi }}.
\end{equation}

On the other hand, for future purposes, we need to know 
the state parameter with respect to the redshift $ z $. 
Taking the assumption that $ {a_0}=1 $ (where $ {a_0}  $ corresponds 
to the value of the scale factor in $ 4D $ at this time), the relation $ 1+z=a_0/a $ leads to
\begin{equation}\label{dt}
\frac{d}{{dt}} = \frac{{da}}{{dt}}\frac{{dz}}{{da}}\frac{d}{{dz}} =  - \left( {1 + z} \right)H\frac{d}{{dz}}.
\end{equation}
Regarding the relation (\ref{dt}), we have
\begin{equation}\label{h2prim}
\dot H =  - \frac{{\left( {1 + z} \right)}}{2}{\left( {{H^2}} \right)^\prime },
\end{equation}

\begin{equation}\label{phii}
\dot \phi  =  - \left( {1 + z} \right)H\phi ',
\end{equation}
and
\begin{equation}\label{phii2}
\ddot \phi  = \left( {1 + z} \right){H^2}\phi ' + \frac{{{{\left( {1 + z} \right)}^2}}}{2}{\left( {{H^2}} \right)^\prime }\phi ' + {\left( {1 + z} \right)^2}{H^2}\phi '',
\end{equation}
where the prime notation represents derivation with respect to the redshift $ z $. 
Inserting the above expressions for $\dot H $, $\dot \phi$ and $\ddot \phi$ into Eq. (\ref{h dot}) and then,
rewriting the relation (\ref{h2prim}) gives
\begin{equation}\label{h22}
{\left( {{H^2}} \right)^\prime } = \left[ {\frac{{4\phi ' + 2\phi ''\left( {1 + z} \right)}}{{\phi  - \phi '\left( {1 + z} \right)}}} \right]{H^2}.
\end{equation}
Integrating the relation (\ref{h22}), the Hubble parameter will be
\begin{equation}\label{hub}
{H^2} = H_0^2\exp \left[ {\int\limits_0^z {\left( {\frac{{4\phi ' + 2\phi ''\left( {1 + z} \right)}}{{\phi - \phi '\left( {1 + z} \right)}}} \right)dz} } \right],
\end{equation}
where $ H_0 $ is the value of the Hubble parameter in the present day (i.e., at $ z=0 $).
Substituting the relation (\ref{h2prim}) into Eq. (\ref{statp}) and using Eq.~(\ref{h22}) gives the state parameter as
\begin{equation}\label{sta}
w(z) =  - 1 + \frac{{4\phi '\left( {1 + z} \right) + 2\phi ''{{\left( {1 + z} \right)}^2}}}{{3\left(\phi  - \phi '\left( {1 + z} \right)\right)}}.
\end{equation}
To obtain the value of the state parameter at the present time,
we assume $ \phi \left( z \right)$ to be a linear function of the redshift as
\begin{equation}\label{phi}
\phi \left( z \right) =  {\phi_ 0} + {\phi '_0}z,
\end{equation}
in which for $ z=0 $, we have $ \phi \left( z \right) =  {\phi_ 0} $.
Using the above assumption in Eq. (\ref{sta}), we reach to the relation
\begin{equation}\label{st2}
w\left( z \right)=  - 1 + \frac{{4{\phi ' _0}\left( {1 + z} \right)}}{ 3\left( {{\phi _0} - {\phi '_0}} \right)},
\end{equation}
for the state parameter. As the value of this parameter in the transition point from the deceleration to the acceleration phase 
of the universe is equal to $ -1/3 $,
hence, from Eq.~(\ref{st2}) we obtain the transition redshift as
\begin{equation}\label{z}
{z_{\rm trans.}} = 
\frac{{{\phi _0}}}{{2{{\phi '}_0}}} - \frac{3}{2}.
\end{equation}
Now, if we consider the value of the transition redshift 
to be equal to its corresponding value in the $ \Lambda CDM $ model, that is $ {z_{\rm trans.}} \simeq 0.67 $, then regarding the Eqs.~(\ref{z}) and (\ref{st2}) the following 
relation is obtained for the state parameter 
\begin{equation}\label{wz}
w\left( z \right) \simeq  - 1 + \frac{{4\left( {1 + z} \right)}}{10.02 }.
\end{equation}
The relation (\ref{wz}) has been plotted in figure 1.
It is illustrated that
at the present time, the value of the state parameter is equal to $ w\left( 0 \right) \simeq  - 0.6 $. As this value is less than $-1/3$, therefore, the existence of the extra dimension can lead to an acceleration phase in the late-time universe.
\begin{figure}[h]
\begin{center}
\includegraphics[scale=0.6]{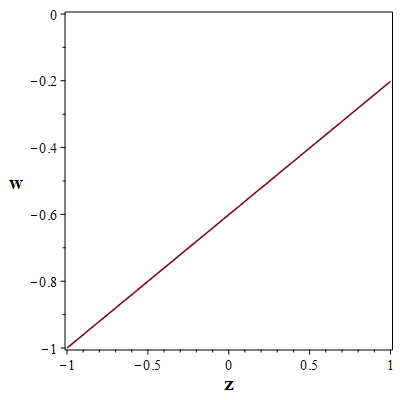}
\caption{The evolution of the state parameter
with respect to the redshift $ z $.}
\end{center}
\end{figure}
\\

\section{Geodesic deviation equation}\label{sec 4}
\indent

As the curvature of the space-time corresponds to the deviation of the parallel geodesics, the geometry of the space-time can also be studied through the geodesic deviation equation (GDE)~\cite{Wald}-\cite{Ellis}. Hence, to complete our investigation of the cosmology of the STM theory, in this section, we derive the geodesic deviation equation associated with this theory. To this purpose, we begin with the general expression for the GDE 
\begin{equation}
\frac{{{D^2}{\eta ^\mu }}}{{D{\nu ^2}}} + ^{(4)}R_{\nu \alpha \beta }^\mu {V^\nu }{\eta ^\alpha }{V^\beta } = 0,
\end{equation}
where $ \nu $ is an affine parameter along the geodesics and $ {\eta ^\mu } $ is the orthogonal deviation vector of two adjacent geodesics. Also, the normalized vector
field $ {V^\mu }={dx^\mu}/d\nu $ is tangent to the geodesics and $ D/D\nu $ is the covariant derivative along the curve.
To obtain the second term in the left-hand side of the above equation, let us notice the well-known expression for the $ 4D $ Riemann tensor
\begin{equation}\label{eq42}
{^{(4)}R_{\mu \nu \alpha \beta}}={^{(4)}C_{\mu \nu \alpha \beta}}
+\frac{1}{2}\left( {{g_{\mu \alpha}}{^{(4)}R_{\nu \beta}} - {g_{\mu
\beta}}{^{(4)}R_{\nu \alpha}} + {g_{\nu \beta}}{^{(4)}R_{\mu \alpha}} -
{g_{\nu \alpha}}{^{(4)}R_{\mu \beta}}} \right) - \frac{1}{6}{^{(4)}R}\left(
{{g_{\mu \alpha}}{g_{\nu \beta}} - {g_{\mu \beta}}{g_{\nu
\alpha}}} \right),
\end{equation}
where in the case of the spatially flat FLRW space-time, the Weyl tensor $ {^{(4)}C_{\mu \nu \alpha \beta}} $ is zero. Hence, by substituting 
the relation (\ref{n5})
into (\ref{eq42}), we obtain
\begin{equation}\label{devi}
 ^{(4)}R_{\nu \alpha \beta }^\mu  {V^\nu }{\eta ^\alpha }{V^\beta }= \left[ {\frac{1}{{2\phi }}\left( {\delta _\alpha ^\mu {\phi _{\nu ;\beta }} - \delta _\beta ^\mu {\phi _{\nu ;\alpha }} + {g_{\nu \beta }}\phi _{;\alpha }^\mu  - {g_{\nu \alpha }}\phi _{;\beta }^\mu } \right) - \frac{^{(4)}R }{{6 }}} \left( {\delta _\alpha ^\mu {g_{\nu \beta }} - \delta _\beta ^\mu {g_{\nu \alpha }}} \right) \right]{V^\nu }{\eta ^\alpha }{V^\beta }.
\end{equation}
Using notations $ \xi  \equiv {V^\mu }{V_\mu } $ and $ E=- {V^0 }{V_0}$ (where $E$ is the total energy), and noticing the point that $ {\eta _\mu }{V^\mu } = 0 $, Eq.~(\ref{devi}) is reduced to
\begin{equation}\label{devii}
^{(4)}R_{\nu \alpha \beta }^\mu {V^\nu }{\eta ^\alpha }{V^\beta }= \frac{1}{2}\left[ {2{E^2}\left( {\frac{{\ddot \phi  - H\dot \phi }}{\phi }} \right) + \xi \left( {\frac{{\ddot \phi  - H\dot \phi }}{{\phi }} + \frac{^{(4)}R}{6}} \right)} \right]{\eta ^\mu }.
\end{equation}
Now, considering (\ref{ro}) and (\ref{press}), Eq.~(\ref{devii}) is rewritten as
\begin{equation}
^{(4)}R_{\nu \alpha \beta }^\mu {V^\nu }{\eta ^\alpha }{V^\beta }=
\frac{1}{2}\left[ {{E^2}\left( {{\rho ^{\left[ {\rm IM} \right]}} + {p^{\left[ {\rm IM} \right]}}} \right) + \frac{\xi }{3}\left( {{\rho ^{\left[ {\rm IM} \right]}} + 3{p^{\left[ {\rm IM} \right]}} + ^{(4)}R} \right)} \right]{\eta ^\mu },
\end{equation}
which is a generalization of the Pirani equation~\cite{Pirani}. Therefore, the GDE of the STM theory for FLRW metric in $ 4D $ will be
\begin{equation}\label{GDE}
\frac{{{D^2}{\eta ^\mu }}}{{D{\nu ^2}}}+\frac{1}{2}\left[ {{E^2}\left( {{\rho ^{\left[ {\rm IM} \right]}} + {p^{\left[ {\rm IM} \right]}}} \right) + \frac{\xi }{3}\left( {{\rho ^{\left[ {\rm IM} \right]}} + 3{p^{\left[ {\rm IM} \right]}} + ^{(4)}R} \right)} \right]{\eta ^\mu }=0.
\end{equation}
Now, let us compare the above GDE with the corresponding ones in the $ \Lambda CDM $ model and in the minimal coupling model of the $ f(R,T) $ modified gravitational theory,
that has been studied in Ref.~\cite{Zare}.
As it is shown in Ref.~\cite{Zare}, the density and the pressure appearing in the GDE of $ \Lambda CDM $ and $ f(R,T) $ are respectively caused by the cosmological constant, and the coupling between the matter and the curvature. However, in the STM theory, the density and the pressure appear in the Eq. (\ref{GDE}) as the effect of the existence of the extra dimension. 
\\
 
\subsection{GDE for time-like vector fields}

In the spatially flat FLRW space-time, the time-like vector fields are determined by the velocities of the comoving observers,
hence we have $ \xi =-1 $, $ E=1 $ and the affine parameter $ \nu $ matches with the proper time $ t $.
Therefore, the GDE for time-like vector fields from Eq.~(\ref{GDE}) reduces to
\begin{equation}\label{gde time}
\frac{{{D^2}{\eta ^\mu }}}{{D{t ^2}}}+\frac{1}{6}\left( {2{\rho ^{\left[ {\rm IM} \right]}} - ^{(4)}R} \right){\eta ^\mu }=0.
\end{equation}
Considering the deviation vector in terms of the comoving frame as $ {\eta ^\mu } = a\left( t \right){e^\mu } $, isotropy implies $ \frac{{D{e^\mu }}}{{Dt}} = 0 $, and
therefore we have
\begin{equation}
\frac{{{D^2}{\eta ^\mu }}}{{D{t^2}}} = \ddot a{e^\mu }.
\end{equation}
According to Eq.~(\ref{tti}), $^{(4)}R=-{^{(4)}T^{[{\rm IM}]}}$. Taking the induced matter as a perfect fluid, i.e., ${^{(4)}T^{[{\rm IM}]}}=-{\rho ^{\left[ {\rm IM} \right]}}+3{p ^{\left[ {\rm IM} \right]}}$, hence, the Eq.~(\ref{gde time}) yields
\begin{equation}
\frac{{\ddot a}}{a} =  - \frac{1}{6}\left( {{\rho ^{\left[ {\rm IM} \right]}} + 3{p^{\left[ {\rm IM} \right]}}} \right).
\end{equation}
This equation is consistent with the Eq.~(\ref{fri2}) in the previous section. The transition point from the deceleration to the acceleration phase of the universe
occurs when $ {\ddot a}=0 $, that corresponds to $ {\rho ^{\left[ {\rm IM} \right]}}=-3 {p ^{\left[ {\rm IM} \right]}}$.
\\

\subsection{GDE for null vector fields}
In this part, we want to investigate the GDE for the past-directed null vector fields in  the STM theory. 
For null vector fields, we have $ \xi=0 $. Considering the
deviation vector in this case as $ {\eta ^\mu} = \eta {e^\mu } $
and using an aligned coordinate base parallel propagated, i.e. ${D{e^\mu }/D\nu } = {V^\alpha }{\nabla_\alpha }{e^\mu } = 0 $, the GDE (\ref{GDE}) reduces to
\begin{equation}\label{nul}
\frac{{{d^2}{\eta }}}{{d{\nu ^2}}} + \frac{E^2}{2}\left( {{\rho ^{\left[ {\rm IM} \right]}} + {p^{\left[ {\rm IM} \right]}}} \right){\eta } = 0.
\end{equation}
It is more suitable to write the deviation vector fields for null geodesics as a function of the redshift to find the observer area-distance. For this purpose, let us start with the following differential operator
\begin{equation}\label{d nu}
\frac{d}{{d\nu }} = \frac{{dz}}{{d\nu }}\frac{d}{{dz}},
\end{equation}
that leads to
\begin{equation}\label{eq43}
\frac{d^2}{d\nu^2}=\left(\frac{dz}{d\nu}\right)^2\frac{d^2}{dz^2}+
\frac{d^2z}{d\nu^2}\frac{d}{dz}=\left(\frac{d\nu}{dz}\right)^{-2}\left[
\frac{d^2}{dz^2} - (\frac{d\nu}{dz})^{-
1}\frac{d^2\nu}{dz^2}\frac{d}{dz}\right].
\end{equation}
In the case of the null geodesics, we have also the relation $(1+z) = a_0/a = E/E_0$, that gives
\begin{equation}\label{dz}
dz =  - \left( {1 + z}\right)\frac{{\dot a}}{a}\frac{{dt}}{{d\nu
}}d\nu  =  - \left( {1 + z} \right)HEd\nu,
\end{equation}
and
\begin{equation}\label{47}
\frac{{{d^2}\nu }}{{d{z^2}}} = \frac{1}{{{E_0}H{{\left( {1 +
z}\right)}^3}}}\left( {2 - \frac{{\dot H}}{{{H^2}}}} \right).
\end{equation}
Now, inserting $ \dot H ={\ddot a}/a - {H^2} $ in the above relation and using Eq. (\ref{fri2}), yields
\begin{equation}\label{48}
\frac{{{d^2}\nu }}{{d{z^2}}} = \frac{1}{{{E_0}H{{\left( {1 +
z}\right)}^3}}}\left[ {3 + \frac{{\left( {{\rho ^{\left[ {\rm IM} \right]}} + 3{p^{\left[
{\rm IM} \right]}}} \right)}}{{6{H^2}}}} \right].
\end{equation}
Then using Eqs. (\ref{eq43}) and (\ref{48}), we obtain
\begin{equation}\label{d2}
\frac{{{d^2}\eta }}{{d{\nu ^2}}} = E_0^2{H^2}{\left( {1 +
z}\right)^4}\left\{ {\eta '' +
\frac{1}{{\left( {1 + z} \right)}}\left[ {3 + \frac{{\left( {{\rho ^{\left[ {\rm IM}
\right]}} + 3{p^{\left[ {\rm IM} \right]}}} \right)}}{{6{H^2}}}}
\right]\eta'} \right\}.
\end{equation}
Finally, substituting Eq.~(\ref{nul}) into (\ref{d2}) gives the null GDE as 
\begin{equation}\label{null}
\eta''+ \frac{3}{{1 + z}}\left[ {1
+\frac{{{1}}}{{18{H^2}}}\left( {{\rho ^{[\rm
IM]}} + 3{p^{[\rm IM]}}} \right)} \right]\eta' +
\frac{{{1}}}{{2{{\left( {1 + z}
\right)}^2}{H^2}}}\left( {{\rho ^{[\rm IM]}} + {p^{[\rm IM]}}}
\right)\eta  = 0.
\end{equation}
Also, inserting Eqs. (\ref{fri1}) and (\ref{statp}) into Eq. (\ref{null}), this equation can be rewritten as the following
\begin{equation}\label{d eta}
\eta''+ \frac{{\left( {7 + 3w} \right)}}{{2\left( {1 + z} \right)}}\eta' + \frac{{3\left( {1 + w} \right)}}{{2{{\left( {1 + z} \right)}^2}}}\eta  = 0.
\end{equation}
Now, by replacing Eq.~(\ref{wz}) in Eq.~(\ref{d eta}), we obtain the GDE for this 
particular model as
\begin{equation}\label{gd}
\eta'' + \frac{{24 + 12.96z}}{{\left( {8.76 + 3.24z} \right)\left( {1 + z} \right)}}\eta' + \frac{{6.48\left( {1 + z} \right)}}{{\left( {8.76 + 3.24z} \right){{\left( {1 + z} \right)}^2}}}\eta  = 0,
\end{equation}
that gives the following solution
\begin{equation}\label{100}
\eta \left( z \right) = \frac{{{C_1}}}{{\left( {73 + 27z} \right)\left( {1 + z} \right)}} + \frac{{{C_2}}}{{73 + 27z}},
\end{equation}
where $ C_1 $ and $ C_2 $ are integration constants which depend on the initial conditions.
We assume that at $ z=0 $, the deviation vector to be zero and its first derivative to be equal to one. We plot
the evolution of the deviation vector fields from Eq. (\ref{100}) in the left diagram of figure 2. In addition, the deviation vector of
the $ \Lambda CDM $ model and the minimal coupling model in $ f(R,T) $ gravity (see
Ref.~\cite{Zare}) has
been plotted in this figure, 
so that the results can be compared with our model.
\begin{figure}[h]
\begin{center}
\includegraphics[scale=0.55]{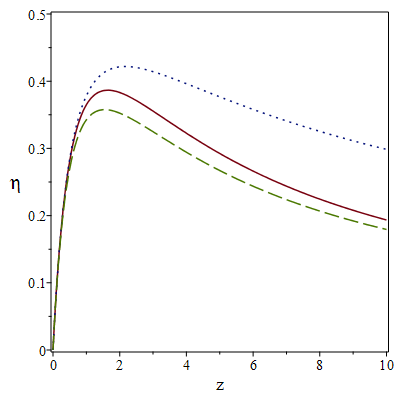}
\includegraphics[scale=0.55]{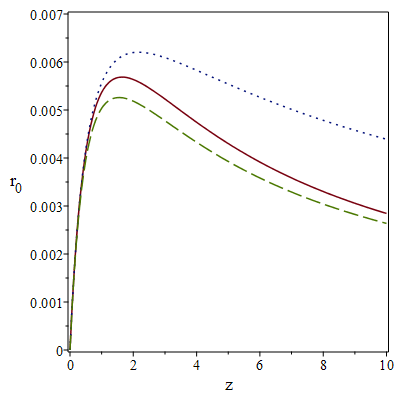}
\caption{The deviation vector (left) and the observer
area-distance (right) are plotted with respect to the redshift
for the particular model in the STM theory (solid curve),
the $\Lambda CDM$ model (dotted curve) and the
$ f(R,T) $ minimal coupling model (dashed curve), with the
appropriate initial conditions $\eta(z)|_{z=0}=0$ and
$\eta'(z)|_{z=0}=1$, and the parametric value $H_0\simeq 68$
(km/s)/Mpc.}
\end{center}
\end{figure}

We also obtain the observer area-distance in the STM theory as a 
measurable quantity.
In FLRW space-time, the magnitude of the deviation vector is related to the proper area $ dA $ (see Refs.~\cite{Mattig}-\cite{Matravers}), and hence the observer area-distance $r_0$, as a function of the redshift $ z $ in units of the present-day Hubble radius $H_0^{- 1} $, is defined as
\begin{equation}\label{r0}
{r_0}\left( z \right) := \sqrt {\left| {\frac{{d{A_0}\left( z
\right)}}{{d\Omega}_0}}\right|}  = \left| {\frac{{\eta \left( {z}
\right)}}{{d\eta \left( {z}
\right)/dl\left| {_{z = 0}} \right.}}} \right|,
\end{equation}
where $\Omega  $ is the solid angle, $ {A_0} $ is the area of the object,
and $ dl = a\left( t \right)dr $ that \textit{r} is the
comoving radial coordinate in the FLRW metric.
By using $d/dl = H\left( {1 + z} \right)d/dz $ and assuming $\eta(0)=0$, we lead to the relation
\begin{equation}\label{eq78}
{r_0}\left( z \right) = \left| {\frac{{\eta \left( z
\right)}}{H(0){d\eta\left( {z} \right)/dz\left| {_{z = 0}}
\right.}}} \right|.
\end{equation}
The plot of Eq.~(\ref{eq78}) for the STM theory has been numerically depicted
in the right diagram of figure 2.
Moreover, the corresponding diagram of two other models in Ref.~\cite{Zare},
has been plotted in this figure.
The comparison between these diagrams demonstrates that, in general, all models have a similar evolution with respect to the redshift.

\section{Conclusion}\label{sec 5}
\indent

In this study, we have considered the space-time-matter theory (STM), as a modified theory of gravity to show how the existence of an extra dimension can lead to a cosmic acceleration phase in the evolution of the universe. To this purpose, at first we have obtained the $ 4D $ Einstein field equations with non-trivial energy-momentum tensor that is induced from the $ 5D $ vacuum space-time. 
In continuation, by considering the generalized spatially flat FLRW metric, we have derived the cosmological equations, and the state parameter as a function of the redshift.
Furthermore, we have shown that assuming the scale factor of the fifth dimension to be a 
linear function of the redshift, 
the value of the state parameter in this particular model will be less that $-1/3$, and hence, the extra dimension can cause an accelerating phase in the 4D universe with no need of the mysterious dark energy.

On the other hand, to make our investigation more instructive, we have derived the geodesic deviation equation in the STM theory for the time-like and null geodesics, to study the relative acceleration of these geodesics as an
effect of the curvature of the space-time.
In the case of the time-like vector fields, we led to the generalized Raychaudhuri equation.
Also, through the geodesic deviation equation of the null vector fields, we have derived the observed area-distance of this model, and have plotted our obtained results with respect to the redshift
to compare our results with the
$ \Lambda CDM $ model and a minimal coupling model of the $ f(R,T) $ gravity.
We have found that the behaviors of the null deviation vector fields, and the observer
area-distance in this specific model of the STM theory, is to a large extent similar to the behaviors of these quantities
in the two mentioned models.
\\

\end{document}